\documentclass[11pt]{article}
\usepackage[margin=2.5cm]{geometry}
\usepackage{amsmath,amssymb,amsthm,bm,mathtools}
\usepackage{authblk}
\usepackage{booktabs}
\usepackage[colorlinks=true,linkcolor=blue,citecolor=blue,urlcolor=cyan]{hyperref}

\newtheorem{proposition}{Proposition}
\newtheorem{corollary}{Corollary}
\theoremstyle{remark}
\newtheorem{remark}{Remark}

\newcommand{\ket}[1]{|#1\rangle}
\newcommand{\bra}[1]{\langle #1|}
\newcommand{\braket}[2]{\langle #1|#2\rangle}
\newcommand{\sech}{\operatorname{sech}}
\newcommand{\I}{\hat I}

\title{What the resolution of the identity knows:\\ exact integral identities from coherent, squeezed,\\ displaced-number and photon-added states}
\author[1]{Ghasem Asadi Cordshooli\thanks{ghascor@iau.ac.ir}}
\author[2]{Mehdi Mirzaee}
\affil[1]{Department of Physics, YI. C., Islamic Azad University, Tehran, Iran}
\affil[2]{Department of Physics, Faculty of Science, Arak University, Arak 38156-8-8349, Iran}
\date{\today}
\begin{document}
\maketitle

\begin{abstract}
A resolution of the identity, projected onto an orthonormal basis, becomes an exact integral identity whose two sides are explicit functions. This paper uses this single step as a common route to a set of results on coherent states and their relatives. For Glauber coherent states the step yields the reproducing property of the Bargmann space and, through its generating function, the orthogonality relations of the number states. Squeezed coherent states yield Gaussian-weighted Hermite identities, whose comparison shows squeezing as a heat flow on the Bargmann space. Displaced number states yield Laguerre identities, and one statement on displaced reference states covers this whole class. Photon-added coherent states resolve the identity on a subspace, and their identity becomes a fractional integral of Riemann--Liouville type. Even and odd coherent states split the identities by parity, and squeezed vacua show where the route ends. Finally, the coherent-state kernel and the Dirac delta both belong to projectors, and orthonormality singles out the delta.
\end{abstract}

\section{Introduction}

Coherent states of the harmonic oscillator form an overcomplete family that resolves the identity~\cite{Glauber1963,Sudarshan1963,KlauderSkagerstam1985}. Textbooks use this resolution mainly as a tool for changing representations and for phase-space and path-integral methods~\cite{KlauderSkagerstam1985,Perelomov1986}. Projected onto an orthonormal basis, the same relation becomes an integral identity with explicit functions on both sides. For Glauber coherent states this identity is the reproducing property of the Bargmann space~\cite{Bargmann1961}, and the coherent-state overlap is a reproducing kernel in the sense of Aronszajn~\cite{Aronszajn1950,Hall2000}.

Related identities for other families appear in separate parts of the literature. In the Bargmann space, squeezed states lead to Hermite polynomials of a complex variable~\cite{Ali2014}, and the Hermite polynomials have a representation through the heat equation~\cite{RosenbloomWidder1959,Widder1975}. Displaced number states, and displaced reference states in general, resolve the identity~\cite{deOliveira1990,Perelomov1986}, a property closely tied to the orthogonality relations of Moyal~\cite{Moyal1949,Grochenig2001}. Photon-added coherent states resolve the identity on a subspace with a Meijer weight~\cite{AgarwalTara1991,SixdeniersPenson2001}, and squeezed vacua admit a resolution after analytic continuation in the Bargmann index~\cite{BrifVourdasMann1996}.

The present paper derives these results from the projected resolution of the identity, Eq.~\eqref{eq:master}, through explicit calculations suited to advanced undergraduate and graduate courses. The same route brings out two connections: comparing two identities for squeezed states reproduces the heat-equation representation of the Hermite polynomials, so that squeezing appears as a heat flow on the Bargmann space, and the identity for photon-added coherent states takes the form of a Riemann--Liouville integral. Table~\ref{tab:summary} collects the families and their identities.

Section~\ref{sec:mechanism} derives the general identity and treats coherent states and the position basis. Sections~\ref{sec:squeezed} to~\ref{sec:cats} treat squeezed coherent states, displaced states, photon-added coherent states, and even and odd coherent states. Section~\ref{sec:boundary} treats the squeezed vacua, and Section~\ref{sec:conclusion} summarizes the results. Exercises close the paper.

\section{The mechanism}\label{sec:mechanism}
\subsection{The general identity}

This subsection states the identity on which the paper rests. Let $\{\ket\gamma\}$ be a family of states that resolves the identity with a positive measure $\mu$,
\begin{equation}\label{eq:roi}
\I=\int d\mu(\gamma)\,\ket\gamma\bra\gamma ,
\end{equation}
and let $\{\ket n\}$ be an orthonormal basis. For every state $\ket\psi$ we have $\I\ket\psi=\ket\psi$, hence
\begin{equation}
\int d\mu(\gamma)\,\ket\gamma\braket{\gamma}{\psi}=\ket\psi ,
\end{equation}
and projection onto $\bra n$ gives
\begin{equation}\label{eq:master}
\int d\mu(\gamma)\,\braket{n}{\gamma}\,\braket{\gamma}{\psi}=\braket{n}{\psi}.
\end{equation}
For the families below, both factors of the integrand and the right-hand side are explicit functions, so Eq.~\eqref{eq:master} is an exact integral identity. Projecting onto a second family of states, in place of $\bra n$, gives identities of the same type.

\subsection{Coherent states and the Bargmann space}

This subsection applies Eq.~\eqref{eq:master} to Glauber coherent states. They read
\begin{equation}\label{eq:cs}
\ket\alpha=e^{-|\alpha|^2/2}\sum_{n\ge0}\frac{\alpha^n}{\sqrt{n!}}\ket n,\qquad \alpha\in\mathbb C,
\end{equation}
so that
\begin{equation}\label{eq:overlap}
\braket{n}{\alpha}=e^{-|\alpha|^2/2}\frac{\alpha^n}{\sqrt{n!}},\qquad
\braket{\beta}{\alpha}=\exp\!\left(-\tfrac12|\alpha|^2-\tfrac12|\beta|^2+\bar\beta\alpha\right),
\end{equation}
and they resolve the identity with $d\mu(\beta)=d^2\beta/\pi$~\cite{Glauber1963}. With $\ket\psi=\ket\alpha$, Eq.~\eqref{eq:master} becomes
\begin{equation}
\int\frac{d^2\beta}{\pi}\,e^{-|\beta|^2/2}\frac{\beta^n}{\sqrt{n!}}\;e^{-|\alpha|^2/2-|\beta|^2/2+\bar\beta\alpha}=e^{-|\alpha|^2/2}\frac{\alpha^n}{\sqrt{n!}} ,
\end{equation}
and division by $e^{-|\alpha|^2/2}/\sqrt{n!}$ gives the master identity
\begin{equation}\label{eq:masterid}
\int d^2\beta\;e^{-|\beta|^2+\bar\beta\alpha}\,\beta^n=\pi\,\alpha^n ,\qquad n=0,1,2,\dots,\quad\alpha\in\mathbb C .
\end{equation}
Linear combinations of Eq.~\eqref{eq:masterid} give the reproducing property of the Bargmann space~\cite{Bargmann1961}.

\begin{proposition}[Reproducing property]\label{prop:bargmann}
For every entire function $f$ with $\int|f(\beta)|^2e^{-|\beta|^2}d^2\beta<\infty$,
\begin{equation}\label{eq:reproducing}
\int\frac{d^2\beta}{\pi}\;e^{-|\beta|^2+\bar\beta\alpha}\,f(\beta)=f(\alpha).
\end{equation}
\end{proposition}

\begin{proof}
The Taylor series $f(\beta)=\sum_nf_n\beta^n$ converges in the weighted norm~\cite{Bargmann1961}, and Eq.~\eqref{eq:masterid} applied term by term gives $\sum_nf_n\alpha^n=f(\alpha)$.
\end{proof}

The exponential in Eq.~\eqref{eq:masterid} is a generating function. With $e^{\bar\beta\alpha}=\sum_m(\bar\beta\alpha)^m/m!$ we have
\begin{equation}
\sum_{m\ge0}\frac{\alpha^m}{m!}\int d^2\beta\;e^{-|\beta|^2}\,\bar\beta^{\,m}\beta^n=\pi\,\alpha^n ,
\end{equation}
and comparing the coefficients of $\alpha^m$ gives the moment relations
\begin{equation}\label{eq:moments}
\int d^2\beta\;e^{-|\beta|^2}\,\bar\beta^{\,m}\beta^n=\pi\,n!\,\delta_{mn}.
\end{equation}
In the Bargmann representation, where $\braket{n}{\beta}\propto\beta^n$, these relations express the orthonormality of the number states. The master identity~\eqref{eq:masterid} is thus the generating function of these orthogonality relations.

\subsection{Projector kernels and the position basis}

This subsection compares the coherent-state kernel with the kernel of the position basis. The position eigenstates resolve the identity, $\I=\int dx\,\ket x\bra x$, and Eq.~\eqref{eq:master} with $\ket n\to\ket{x'}$ and $\ket\psi=\ket{x_0}$ reads
\begin{equation}\label{eq:projx}
\int dx\;\braket{x'}{x}\,\braket{x}{x_0}=\braket{x'}{x_0}.
\end{equation}
The coherent states give the same relation,
\begin{equation}\label{eq:projcs}
\int\frac{d^2\beta}{\pi}\;\braket{\alpha'}{\beta}\,\braket{\beta}{\alpha}=\braket{\alpha'}{\alpha}.
\end{equation}
Both equations state that the kernel belongs to a projector, as for every reproducing kernel~\cite{Aronszajn1950}. Indeed, the resolution of the identity makes the map $V:\ket\psi\mapsto\psi(\gamma)=\braket{\gamma}{\psi}$ an isometry into $L^2(d\mu)$, the square-integrable functions on the label space. The operator $VV^\dagger$ has the kernel $\braket{\gamma}{\gamma'}$ and projects onto the range of $V$. For the position basis $V$ maps onto all of $L^2(dx)$, so $VV^\dagger=1$ and
\begin{equation}
\braket{x}{x_0}=\delta(x-x_0),
\end{equation}
with the delta function understood in the distributional sense that a continuous basis requires. For coherent states the range of $V$ consists of the functions $e^{-|\beta|^2/2}f(\bar\beta)$ with $f$ entire, a proper subspace, and the projector kernel is the smooth Gaussian~\eqref{eq:overlap}. Orthonormality of the family thus selects the delta among the projector kernels.

\begin{remark}[Semiclassical limit]
For an oscillator with Planck constant $\hbar$ the coherent states obey
\begin{equation}
\braket{\beta}{\alpha}_\hbar=\exp\!\left[\frac{2\bar\beta\alpha-|\alpha|^2-|\beta|^2}{2\hbar}\right],\qquad \I=\int\frac{d^2\beta}{\pi\hbar}\,\ket\beta\bra\beta ,
\end{equation}
and the resolution of the identity holds for every $\hbar>0$. The probability density
\begin{equation}
\frac{|\braket{\beta}{\alpha}_\hbar|^2}{\pi\hbar}=\frac{1}{\pi\hbar}\,e^{-|\beta-\alpha|^2/\hbar}
\end{equation}
has unit integral and tends to $\delta^{(2)}(\beta-\alpha)$ as $\hbar\to0$: the semiclassical limit localizes the coherent states, while each resolution of the identity stays exact.
\end{remark}

\section{Squeezed coherent states and heat flow}\label{sec:squeezed}

This section applies Eq.~\eqref{eq:master} to squeezed coherent states and relates squeezing to a heat flow.

\subsection{Resolution of the identity and amplitudes}

This subsection collects the properties of squeezed coherent states used below. With the squeezing operator $\hat S(\xi)=\exp[\frac12(\bar\xi\hat a^2-\xi\hat a^{\dagger2})]$, $\xi=r\,e^{i\vartheta}$, and the displacement operator $\hat D(\beta)$, the squeezed coherent states read $\ket{\beta,\xi}=\hat D(\beta)\hat S(\xi)\ket0$~\cite{Yuen1976}. The relation $\hat S^\dagger\hat a\hat S=\hat a\cosh r-\hat a^\dagger e^{i\vartheta}\sinh r$ gives
\begin{equation}\label{eq:DS}
\hat D(\beta)\hat S(\xi)=\hat S(\xi)\hat D(u),\qquad u=\beta\cosh r+\bar\beta\,e^{i\vartheta}\sinh r .
\end{equation}
The map $\beta\mapsto u$ has the Jacobian $\cosh^2r-\sinh^2r=1$, hence
\begin{equation}
\int\frac{d^2\beta}{\pi}\,\ket{\beta,\xi}\bra{\beta,\xi}=\hat S\left(\int\frac{d^2u}{\pi}\,\ket u\bra u\right)\hat S^\dagger=\I .
\end{equation}
The number-state amplitudes read~\cite{Yuen1976,GerryKnight2005}
\begin{equation}\label{eq:sqamp}
\braket{n}{\beta,\xi}=\frac{e^{-|u|^2/2+cu^2}}{\sqrt{\cosh r}}\,\frac{h_n(u)}{\sqrt{n!}},\qquad c=\tfrac12e^{-i\vartheta}\tanh r ,
\end{equation}
with the polynomials
\begin{equation}\label{eq:hn}
h_n(u)=\left(\tfrac12e^{i\vartheta}\tanh r\right)^{n/2}H_n\!\left(\frac{u}{\sqrt{e^{i\vartheta}\sinh2r}}\right)=\sum_{k}\frac{n!}{k!\,(n-2k)!}\left(-\tfrac12e^{i\vartheta}\tanh r\right)^k\left(\frac{u}{\cosh r}\right)^{n-2k},
\end{equation}
where $H_n$ are the Hermite polynomials. At $r=0$ we have $h_n(u)=u^n$. In the Bargmann space these amplitudes are Hermite polynomials of a complex variable times a Gaussian factor~\cite{Ali2014}.

\subsection{Squeezed target}

This subsection takes a squeezed coherent state as the target. With $\ket\psi=\ket{\alpha,\xi}$ and $v=\alpha\cosh r+\bar\alpha\,e^{i\vartheta}\sinh r$, Eq.~\eqref{eq:DS} gives $\braket{\beta,\xi}{\alpha,\xi}=\braket{u}{v}$, and Eq.~\eqref{eq:master} becomes
\begin{equation}
\int\frac{d^2u}{\pi}\;\frac{e^{-|u|^2/2+cu^2}h_n(u)}{\sqrt{\cosh r\,n!}}\;e^{-|u|^2/2-|v|^2/2+\bar uv}=\frac{e^{-|v|^2/2+cv^2}h_n(v)}{\sqrt{\cosh r\,n!}} .
\end{equation}

\begin{proposition}
For every $v\in\mathbb C$ and $n\ge0$,
\begin{equation}\label{eq:sqsq}
\int d^2u\;e^{-|u|^2+\bar uv+cu^2}\,h_n(u)=\pi\,e^{cv^2}\,h_n(v).
\end{equation}
\end{proposition}

Equation~\eqref{eq:sqsq} is the reproducing property~\eqref{eq:reproducing} for the entire function $e^{cu^2}h_n(u)$, which belongs to the Bargmann space because $|c|=\frac12\tanh r<\frac12$.

\subsection{Coherent target}

This subsection takes a Glauber coherent state as the target. With $\ket\psi=\ket\alpha$ we need the mixed overlap. Equation~\eqref{eq:DS} and $\hat S^\dagger\ket\alpha=\hat D(v)\hat S(-\xi)\ket0$ give
\begin{equation}
\braket{\beta,\xi}{\alpha}=\frac{1}{\sqrt{\cosh r}}\exp\!\left[-\tfrac12|u|^2-\tfrac12|v|^2+\bar uv+\bar c\,(\bar u-\bar v)^2\right].
\end{equation}
In the product with Eq.~\eqref{eq:sqamp}, two simplifications occur:
\begin{equation}
v-e^{i\vartheta}\tanh r\;\bar v=\frac{\alpha}{\cosh r},\qquad
\tfrac12|v|^2-\tfrac12|\alpha|^2-\bar c\,\bar v^{\,2}=c\,\alpha^2 .
\end{equation}

\begin{proposition}
For every $\alpha\in\mathbb C$ and $n\ge0$,
\begin{equation}\label{eq:sqcoh}
\int d^2u\;\exp\!\left[-|u|^2+\frac{\bar u\,\alpha}{\cosh r}+c\,u^2+\bar c\,\bar u^{\,2}\right]h_n(u)=\pi\cosh r\;e^{c\alpha^2}\,\alpha^n .
\end{equation}
\end{proposition}

The integrand of Eq.~\eqref{eq:sqcoh} contains $\bar u^{\,2}$, so this identity goes beyond the reproducing property, while its right-hand side keeps the monomial $\alpha^n$ of the master identity. At $n=0$ the identity is the standard two-dimensional Gaussian integral with determinant $1-\tanh^2r=\sech^2r$, which produces the factor $\cosh r$.

\subsection{Squeezing as a heat flow}

This subsection relates the two identities. Since $\bar u\,e^{\bar ua}=\partial_ae^{\bar ua}$, the factor $e^{\bar c\bar u^2}$ acts as the operator $e^{\bar c\,\partial_a^2}$ on $e^{\bar ua}$. With $a=\alpha/\cosh r$ and Proposition~\ref{prop:bargmann}, the left-hand side of Eq.~\eqref{eq:sqcoh} becomes
\begin{equation}
e^{\bar c\,\partial_a^2}\int d^2u\;e^{-|u|^2+\bar ua}\,e^{cu^2}h_n(u)=\pi\,e^{\bar c\,\partial_a^2}\Big[e^{ca^2}h_n(a)\Big].
\end{equation}
Hence Eq.~\eqref{eq:sqcoh} is equivalent to
\begin{equation}\label{eq:heatform}
e^{\bar c\,\partial_a^2}\Big[e^{ca^2}h_n(a)\Big]=\cosh r\;e^{c\cosh^2r\,a^2}\,\big(a\cosh r\big)^n .
\end{equation}
The Gaussian integral representation of the heat semigroup gives, for $|4\lambda c|<1$,
\begin{equation}
e^{\lambda\partial_a^2}\Big[e^{ca^2}f(a)\Big]=\frac{1}{\sqrt{1-4\lambda c}}\;\exp\!\left(\frac{ca^2}{1-4\lambda c}\right)\Big[e^{\lambda'\partial_y^2}f\Big]\!\left(\frac{a}{1-4\lambda c}\right),\qquad \lambda'=\frac{\lambda}{1-4\lambda c}.
\end{equation}
For $\lambda=\bar c$ one has $4\lambda c=\tanh^2r$, $1-4\lambda c=\sech^2r$ and $\lambda'=\bar c\cosh^2r$. With $y=x\sqrt{e^{i\vartheta}\sinh2r}$ in Eq.~\eqref{eq:hn},
\begin{equation}
\frac{\lambda'}{e^{i\vartheta}\sinh2r}=\frac{\frac12\tanh r\,\cosh^2r}{2\sinh r\cosh r}=\frac14 ,
\end{equation}
while the prefactor in Eq.~\eqref{eq:hn} turns $(2x)^n$ into $(y/\cosh r)^n$, which equals $(a\cosh r)^n$ at $y=a\cosh^2r$. Equation~\eqref{eq:heatform} thus reduces to the following statement.

\begin{proposition}[Squeezing as heat flow]\label{prop:heat}
Given Proposition~\ref{prop:bargmann}, the identity~\eqref{eq:sqcoh} is equivalent to
\begin{equation}\label{eq:hermiteheat}
e^{\frac14\frac{d^2}{dx^2}}\,H_n(x)=(2x)^n ,\qquad n=0,1,2,\dots
\end{equation}
\end{proposition}

Equation~\eqref{eq:hermiteheat} is the heat-polynomial representation of the Hermite polynomials~\cite{RosenbloomWidder1959,Widder1975}: evolved by the heat equation for the time one quarter, each Hermite polynomial becomes a monomial. In the Bargmann representation the squeezing operator acts through Gaussian integral kernels~\cite{Folland1989}, and Proposition~\ref{prop:heat} isolates the heat flow inside this action.

\section{Displaced number states and displaced reference states}\label{sec:displaced}

This section treats displaced number states and then displaced reference states of arbitrary form.

\subsection{Displaced number states}

This subsection derives a Laguerre identity. The displaced number states $\hat D(\beta)\ket m$~\cite{deOliveira1990} resolve the identity with $d^2\beta/\pi$ for every fixed $m$ (Proposition~\ref{prop:fiducial} below). With $\hat D^\dagger(\beta)\ket\alpha=e^{(\bar\beta\alpha-\beta\bar\alpha)/2}\ket{\alpha-\beta}$ we have
\begin{equation}
\bra m\hat D^\dagger(\beta)\ket\alpha=e^{(\bar\beta\alpha-\beta\bar\alpha)/2}\,e^{-|\alpha-\beta|^2/2}\frac{(\alpha-\beta)^m}{\sqrt{m!}}=\braket{\beta}{\alpha}\,\frac{(\alpha-\beta)^m}{\sqrt{m!}} ,
\end{equation}
and for $n\ge m$ the matrix elements read~\cite{CahillGlauber1969}
\begin{equation}
\bra n\hat D(\beta)\ket m=\sqrt{\frac{m!}{n!}}\;\beta^{\,n-m}\,e^{-|\beta|^2/2}\,L_m^{(n-m)}\big(|\beta|^2\big),
\end{equation}
with the associated Laguerre polynomials $L_m^{(n-m)}$. Equation~\eqref{eq:master} with $\ket\psi=\ket\alpha$ then gives the following identity.

\begin{proposition}
For $n\ge m\ge0$ and every $\alpha\in\mathbb C$,
\begin{equation}\label{eq:laguerre}
\int d^2\beta\;e^{-|\beta|^2+\bar\beta\alpha}\,\beta^{\,n-m}\,(\alpha-\beta)^m\,L_m^{(n-m)}\big(|\beta|^2\big)=\pi\,\alpha^n .
\end{equation}
\end{proposition}

At $m=0$ Eq.~\eqref{eq:laguerre} reduces to the master identity~\eqref{eq:masterid}. For $m\ge1$ the integrand depends on $|\beta|^2$ through the Laguerre polynomial, while the right-hand side keeps the monomial $\alpha^n$.

\subsection{Displaced reference states}

This subsection extends the construction to an arbitrary reference state $\ket\phi$. We define the Bargmann function
\begin{equation}
F_\phi(\gamma)=e^{|\gamma|^2/2}\braket{\phi}{\gamma}=\sum_{k\ge0}\braket{\phi}{k}\frac{\gamma^k}{\sqrt{k!}} ,
\end{equation}
an entire function of $\gamma$.

\begin{proposition}\label{prop:fiducial}
For every normalized state $\ket\phi$,
\begin{equation}\label{eq:fiducial}
\int\frac{d^2\beta}{\pi}\,\hat D(\beta)\ket\phi\bra\phi\hat D^\dagger(\beta)=\I,\qquad
\bra\phi\hat D^\dagger(\beta)\ket\alpha=\braket{\beta}{\alpha}\,F_\phi(\alpha-\beta).
\end{equation}
\end{proposition}

\begin{proof}
Let $\hat A$ denote the integral. The relation $\hat D(\gamma)\hat D(\beta)=e^{i\,\mathrm{Im}(\gamma\bar\beta)}\hat D(\gamma+\beta)$ and the translation invariance of $d^2\beta$ give $\hat D(\gamma)\hat A\hat D^\dagger(\gamma)=\hat A$ for every $\gamma$. The displacement operators act irreducibly, so $\hat A=a\,\I$ by Schur's lemma, and the coherent-state resolution of the identity gives
\begin{equation}
a=\bra0\hat A\ket0=\int\frac{d^2\beta}{\pi}\,\big|\braket{-\beta}{\phi}\big|^2=\bra\phi\left(\int\frac{d^2\beta}{\pi}\,\ket{-\beta}\bra{-\beta}\right)\ket\phi=1 .
\end{equation}
For the overlap,
\begin{equation}
\bra\phi\hat D^\dagger(\beta)\ket\alpha=e^{(\bar\beta\alpha-\beta\bar\alpha)/2}\braket{\phi}{\alpha-\beta}=e^{(\bar\beta\alpha-\beta\bar\alpha)/2}\,e^{-|\alpha-\beta|^2/2}F_\phi(\alpha-\beta)=\braket{\beta}{\alpha}F_\phi(\alpha-\beta).
\end{equation}
\end{proof}

The resolution in Eq.~\eqref{eq:fiducial} is equivalent to the orthogonality relations of Moyal for the displacement operators~\cite{Moyal1949,Grochenig2001}. Projecting Eq.~\eqref{eq:fiducial} onto a coherent state $\bra z$, in place of $\bra n$, gives a single identity whose Taylor coefficients in $\bar z$ are the number-state projections.

\begin{corollary}
For every normalized $\ket\phi$ and $z,\alpha\in\mathbb C$,
\begin{equation}\label{eq:twosided}
\int d^2\beta\;e^{-|\beta|^2+\bar z\beta+\bar\beta\alpha}\;\overline{F_\phi(z-\beta)}\;F_\phi(\alpha-\beta)=\pi\,e^{\bar z\alpha}.
\end{equation}
\end{corollary}

\begin{proof}
Proposition~\ref{prop:fiducial} gives $\bra z\hat D(\beta)\ket\phi=\braket{z}{\beta}\overline{F_\phi(z-\beta)}$, and
\begin{equation}
\braket{z}{\beta}\braket{\beta}{\alpha}=e^{-|z|^2/2-|\alpha|^2/2}\,e^{-|\beta|^2+\bar z\beta+\bar\beta\alpha},\qquad \braket{z}{\alpha}=e^{-|z|^2/2-|\alpha|^2/2}\,e^{\bar z\alpha}.
\end{equation}
\end{proof}

The Bargmann functions of photon-added and photon-subtracted reference states follow from $\hat a\ket\gamma=\gamma\ket\gamma$ and $\hat a^\dagger\ket\gamma=(\partial_\gamma+\frac12\bar\gamma)\ket\gamma$:
\begin{equation}\label{eq:addsub}
F_{\hat a^\dagger\phi}(\gamma)=\gamma\,F_\phi(\gamma),\qquad F_{\hat a\phi}(\gamma)=\partial_\gamma F_\phi(\gamma).
\end{equation}
Adding a photon thus multiplies the Bargmann function by $\gamma$, and subtracting a photon differentiates it. Three examples illustrate Eq.~\eqref{eq:twosided}:
\begin{itemize}
\item the number state $\ket m=(\hat a^\dagger)^m\ket0/\sqrt{m!}$ gives $F(\gamma)=\gamma^m/\sqrt{m!}$, and Eq.~\eqref{eq:twosided} contains Eq.~\eqref{eq:laguerre};
\item the squeezed vacuum $\hat S(\xi)\ket0$ gives $F(\gamma)=(\cosh r)^{-1/2}e^{-c\gamma^2}$, and Eq.~\eqref{eq:twosided} contains Eqs.~\eqref{eq:sqsq} and~\eqref{eq:sqcoh};
\item the photon-subtracted squeezed vacuum obeys $\hat a\hat S(\xi)\ket0=-e^{i\vartheta}\sinh r\,\hat S(\xi)\ket1$, and Eq.~\eqref{eq:addsub} gives $F(\gamma)=(\cosh r)^{-3/2}\,\gamma\,e^{-c\gamma^2}$ for the normalized state $\hat S(\xi)\ket1$, hence
\begin{equation}
\int d^2\beta\;e^{-|\beta|^2+\bar z\beta+\bar\beta\alpha}\,(\bar z-\bar\beta)(\alpha-\beta)\,e^{-\bar c(\bar z-\bar\beta)^2-c(\alpha-\beta)^2}=\pi\cosh^3r\;e^{\bar z\alpha}.
\end{equation}
\end{itemize}

\section{Photon-added coherent states and fractional integration}\label{sec:added}

This section treats the photon-added coherent states $\hat a^{\dagger m}\ket\beta$~\cite{AgarwalTara1991}, which yield a fractional integration in place of the reproducing property.

\subsection{Resolution of the identity on a subspace}

This subsection constructs the resolving measure. The states $\hat a^{\dagger m}\ket\beta$ have vanishing components on $\ket0,\dots,\ket{m-1}$, so they resolve at most the projector
\begin{equation}
\hat P_m=\sum_{k\ge m}\ket k\bra k .
\end{equation}
We use the function
\begin{equation}\label{eq:Gm}
G_m(x)=\frac{1}{(m-1)!}\int_0^1\frac{dt}{t}\,(1-t)^{m-1}e^{-x/t},\qquad m\ge1,
\end{equation}
which is positive, coincides with the exponential integral $E_1(x)$ at $m=1$, and has the Mellin transform
\begin{equation}\label{eq:mellin}
\int_0^\infty x^{s-1}G_m(x)\,dx=\frac{\Gamma(s)}{(m-1)!}\int_0^1t^{s-1}(1-t)^{m-1}dt=\frac{\Gamma(s)^2}{\Gamma(s+m)} ,
\end{equation}
so that $G_m$ is the Meijer function $G^{2,0}_{1,2}\big(x\,\big|\,{m\atop 0,0}\big)$~\cite{DLMF}.

\begin{proposition}\label{prop:added}
For $m\ge1$,
\begin{equation}\label{eq:roiadded}
\int\frac{d^2\beta}{\pi}\;e^{|\beta|^2}G_m\big(|\beta|^2\big)\;\hat a^{\dagger m}\ket\beta\bra\beta\hat a^m=\hat P_m .
\end{equation}
\end{proposition}

\begin{proof}
For $k\ge m$ we have $\bra k\hat a^{\dagger m}\ket\beta=e^{-|\beta|^2/2}\beta^{k-m}\sqrt{k!}/(k-m)!$. The angular integral makes the left-hand side diagonal, and with $x=|\beta|^2$ and $j=k-m$ its diagonal elements read
\begin{equation}
\int_0^\infty dx\;G_m(x)\,x^j\,\frac{(j+m)!}{(j!)^2}=\frac{\Gamma(j+1)^2}{\Gamma(j+1+m)}\,\frac{(j+m)!}{(j!)^2}=1 ,
\end{equation}
where Eq.~\eqref{eq:mellin} at $s=j+1$ was used.
\end{proof}

This resolution of the identity appears in Ref.~\cite{SixdeniersPenson2001}, and the identities below follow from it.

\subsection{Identities}

This subsection projects Eq.~\eqref{eq:roiadded} with a coherent target. With $\hat a^m\ket\alpha=\alpha^m\ket\alpha$ we have $\bra\beta\hat a^m\ket\alpha=\alpha^m\braket{\beta}{\alpha}$, and projecting Eq.~\eqref{eq:roiadded} onto $\bra n$, with $n=m+j$, gives
\begin{equation}
\alpha^m\,e^{-|\alpha|^2/2}\frac{\sqrt{n!}}{j!}\int\frac{d^2\beta}{\pi}\,G_m\big(|\beta|^2\big)\,e^{\bar\beta\alpha}\,\beta^{\,j}=e^{-|\alpha|^2/2}\frac{\alpha^n}{\sqrt{n!}} .
\end{equation}

\begin{proposition}\label{prop:rl}
For $j\ge0$, $m\ge1$ and every $\alpha\in\mathbb C$,
\begin{equation}\label{eq:addedid}
\int d^2\beta\;G_m\big(|\beta|^2\big)\,e^{\bar\beta\alpha}\,\beta^{\,j}=\pi\,\frac{j!}{(j+m)!}\,\alpha^{j}.
\end{equation}
For every entire function $f$ of order below two,
\begin{equation}\label{eq:RL}
\int d^2\beta\;G_m\big(|\beta|^2\big)\,e^{\bar\beta\alpha}\,f(\beta)=\frac{\pi}{(m-1)!}\int_0^1(1-s)^{m-1}f(s\alpha)\,ds=\frac{\pi}{\alpha^m}\,\big(I^mf\big)(\alpha),
\end{equation}
where $(I^mf)(\alpha)=\frac{1}{(m-1)!}\int_0^\alpha(\alpha-t)^{m-1}f(t)\,dt$ is the Riemann--Liouville integral of order $m$~\cite{Samko1993}.
\end{proposition}

\begin{proof}
Equation~\eqref{eq:addedid} follows from the projection above. For $f(\beta)=\beta^j$,
\begin{equation}
\frac{1}{(m-1)!}\int_0^1(1-s)^{m-1}s^j\,ds=\frac{B(j+1,m)}{(m-1)!}=\frac{j!}{(j+m)!},
\end{equation}
which matches Eq.~\eqref{eq:addedid}, and linearity extends the identity to Taylor series. The substitution $t=s\alpha$ gives the Riemann--Liouville form.
\end{proof}

The coherent states reproduce a function at the point $\alpha$, Eq.~\eqref{eq:reproducing}, while the photon-added coherent states average it along the segment from $0$ to $\alpha$ with the weight $(1-s)^{m-1}$. At $m=1$, Eq.~\eqref{eq:addedid} reads
\begin{equation}
\int d^2\beta\;E_1\big(|\beta|^2\big)\,e^{\bar\beta\alpha}\,\beta^{\,j}=\frac{\pi\,\alpha^j}{j+1}.
\end{equation}
The photon-subtracted coherent states $\hat a^m\ket\beta=\beta^m\ket\beta$ differ from $\ket\beta$ by a phase after normalization, so they reproduce the coherent-state identities.

\section{Superpositions of coherent states}\label{sec:cats}

This section treats superpositions of coherent states, which split the master identity by parity. The even and odd coherent states~\cite{Dodonov1974}
\begin{equation}
\ket\beta_\pm=\mathcal N_\pm\big(\ket\beta\pm\ket{-\beta}\big),\qquad \mathcal N_\pm^{-2}=2\left(1\pm e^{-2|\beta|^2}\right),
\end{equation}
obey $\ket\beta_\pm\propto(1\pm\hat\Pi)\ket\beta$ with the parity operator $\hat\Pi=e^{i\pi\hat n}$. The coherent-state resolution of the identity gives
\begin{equation}
\int\frac{d^2\beta}{2\pi}\left(1\pm e^{-2|\beta|^2}\right)\ket\beta_\pm\bra\beta_\pm=\frac{1\pm\hat\Pi}{2},
\end{equation}
so each parity family resolves the projector onto its parity sector. The overlaps read
\begin{equation}
{}_\pm\!\braket{\beta}{\alpha}=2\,\mathcal N_\pm\,e^{-|\alpha|^2/2-|\beta|^2/2}\times\begin{cases}\cosh(\bar\beta\alpha)\\ \sinh(\bar\beta\alpha)\end{cases},
\end{equation}
the weight $1\pm e^{-2|\beta|^2}$ cancels the normalization, and Eq.~\eqref{eq:master} gives
\begin{equation}\label{eq:cat}
\int d^2\beta\;e^{-|\beta|^2}\cosh(\bar\beta\alpha)\,\beta^n=\pi\,\alpha^n\,\frac{1+(-1)^n}{2},\qquad
\int d^2\beta\;e^{-|\beta|^2}\sinh(\bar\beta\alpha)\,\beta^n=\pi\,\alpha^n\,\frac{1-(-1)^n}{2}.
\end{equation}
The sum of the two identities is the master identity~\eqref{eq:masterid}. Superpositions of the $N$ coherent states $\ket{\omega^k\beta}$, with $\omega=e^{2\pi i/N}$, split it into $N$ residue classes,
\begin{equation}
\int d^2\beta\;e^{-|\beta|^2}E_{N,j}\big(\bar\beta\alpha\big)\,\beta^n=\pi\,\alpha^n\,\delta_{n\equiv j\ (\mathrm{mod}\ N)},\qquad E_{N,j}(x)=\sum_{k\equiv j\ (\mathrm{mod}\ N)}\frac{x^k}{k!},
\end{equation}
with the generalized hyperbolic functions $E_{N,j}$, which reduce to $\cosh$ and $\sinh$ at $N=2$.

\section{The boundary of the method: squeezed vacua}\label{sec:boundary}

This section shows that spanning a subspace is a weaker property than resolving its identity with a positive measure. The squeezed vacua $\hat S(\xi)\ket0$ form such a family. With $\zeta=-e^{i\vartheta}\tanh r$ in the unit disc they read~\cite{GerryKnight2005}
\begin{equation}
\hat S(\xi)\ket0=\left(1-|\zeta|^2\right)^{1/4}\sum_{m\ge0}\frac{\sqrt{(2m)!}}{2^m\,m!}\,\zeta^m\ket{2m},
\end{equation}
so they span the even subspace. Their components obey
\begin{equation}
\big|\braket{2m}{\zeta}\big|^2=c_m\left(1-|\zeta|^2\right)^{1/2}|\zeta|^{2m},\qquad c_m=\frac{(2m)!}{4^m\,(m!)^2},\qquad \frac{c_{m+1}}{c_m}=\frac{2m+1}{2m+2}.
\end{equation}

\begin{proposition}
For every positive measure $\mu$ on the unit disc, the operator $\hat A=\int d\mu(\zeta)\,\ket\zeta\bra\zeta$ built from the squeezed vacua differs from the projector onto the even subspace.
\end{proposition}

\begin{proof}
Let $a_m=\int d\mu(\zeta)\,(1-|\zeta|^2)^{1/2}|\zeta|^{2m}$. Since $|\zeta|<1$, we have $a_{m+1}\le a_m$. A resolution of the even projector requires
\begin{equation}
\bra{2m}\hat A\ket{2m}=c_m\,a_m=1\qquad\text{for all }m,
\end{equation}
hence $a_m=1/c_m$ and
\begin{equation}
\frac{a_{m+1}}{a_m}=\frac{c_m}{c_{m+1}}=\frac{2m+2}{2m+1}>1,
\end{equation}
which contradicts $a_{m+1}\le a_m$.
\end{proof}

A signed resolution of the identity, obtained by analytic continuation in the Bargmann index of the underlying SU(1,1) coherent states, exists for this family~\cite{BrifVourdasMann1996}. Within the positive-measure framework of Eq.~\eqref{eq:roi}, the squeezed vacua thus mark the boundary of the method.

\section{Conclusion}\label{sec:conclusion}

This section summarizes the results. Projecting the resolution of the identity onto a basis gives an exact integral identity, Eq.~\eqref{eq:master}. For Glauber coherent states this identity is the reproducing property of the Bargmann space and, read as a generating function, it yields the orthogonality relations of the number states. Applied to the other families, the same step recovers the identities collected in Table~\ref{tab:summary}: Gaussian-weighted Hermite identities, Laguerre identities, a two-sided identity for every displaced reference state, a parity splitting for even and odd coherent states, and the boundary set by the squeezed vacua. Two connections emerge along the way: squeezing acts on the Bargmann space as a heat flow, and photon addition turns the reproducing property into a Riemann--Liouville integration. The comparison with the position basis shows that both kernels belong to projectors and that orthonormality singles out the Dirac delta.

\begin{table}[t]
\centering
\caption{Families of states, their resolving measures, and the identities they generate.}
\label{tab:summary}
\footnotesize
\begin{tabular}{lll}
\toprule
family & resolving measure & identity\\
\midrule
coherent $\ket\beta$ & $d^2\beta/\pi$ & reproducing, Eqs.~\eqref{eq:masterid}--\eqref{eq:moments}\\
squeezed coherent $\ket{\beta,\xi}$ & $d^2\beta/\pi$ & Hermite--Gaussian, heat flow, Eqs.~\eqref{eq:sqsq}--\eqref{eq:hermiteheat}\\
displaced number $\hat D(\beta)\ket m$ & $d^2\beta/\pi$ & Laguerre, Eq.~\eqref{eq:laguerre}\\
displaced reference $\hat D(\beta)\ket\phi$ & $d^2\beta/\pi$ & two-sided identity, Eq.~\eqref{eq:twosided}\\
photon-added $\hat a^{\dagger m}\ket\beta$ & $e^{|\beta|^2}G_m\,d^2\beta/\pi$, on $\hat P_m$ & Riemann--Liouville, Eq.~\eqref{eq:RL}\\
even and odd $\ket\beta_\pm$ & $(1\pm e^{-2|\beta|^2})\,d^2\beta/2\pi$, on $\frac{1\pm\hat\Pi}{2}$ & parity split, Eq.~\eqref{eq:cat}\\
squeezed vacua $\hat S(\xi)\ket0$ & positive measures excluded & moments, Sec.~\ref{sec:boundary}\\
\bottomrule
\end{tabular}
\end{table}

\section*{Exercises}

\textbf{Exercise 1.} Derive the moment relations~\eqref{eq:moments} for $(m,n)=(0,0)$, $(1,1)$ and $(1,0)$ directly in polar coordinates, and compare with the generating-function derivation of Section~\ref{sec:mechanism}.

\textbf{Exercise 2.} Write $\beta=x+iy$ and show that Eq.~\eqref{eq:masterid} with $n=1$ and real $\alpha$ follows from the one-dimensional integrals $\int_{-\infty}^\infty x\,e^{-x^2+\alpha x}\,dx=\frac{\sqrt\pi}{2}\,\alpha\,e^{\alpha^2/4}$ and $\int_{-\infty}^\infty e^{-y^2-i\alpha y}\,dy=\sqrt\pi\,e^{-\alpha^2/4}$.

\textbf{Exercise 3.} Verify Eq.~\eqref{eq:laguerre} for $n=m=1$, where $L_1^{(0)}(x)=1-x$, by expanding the integrand and using Eq.~\eqref{eq:moments}.

\textbf{Exercise 4.} Verify Eq.~\eqref{eq:hermiteheat} for $n=2$ and $n=3$ by expanding the exponential operator, and evaluate Eq.~\eqref{eq:sqcoh} at $n=0$ as a two-dimensional Gaussian integral.

\textbf{Exercise 5.} Prove Eq.~\eqref{eq:addsub} and use it to find the Bargmann function of the photon-added squeezed vacuum $\hat a^\dagger\hat S(\xi)\ket0$.

\textbf{Exercise 6.} Show that $G_1(x)=E_1(x)$, compute $\int d^2\beta\,E_1(|\beta|^2)\,e^{\bar\beta\alpha}e^{\kappa\beta}$ with Eq.~\eqref{eq:RL}, and check the result against the series of Eq.~\eqref{eq:addedid}.

\textbf{Exercise 7.} Show that the squeezed vacua give $a_0=1$ and $a_1=2$ in the notation of Section~\ref{sec:boundary}, and explain why this pair of values already excludes a positive resolving measure.

\end{document}